\documentclass[12pt]{article}
\usepackage{graphicx}
\begin{document}

\centerline{\bf \large Sociophysics Simulations III: Retirement
Demography}

\bigskip
\centerline{Lotfi Zekri* and Dietrich Stauffer}

\bigskip
\centerline{Institute for Theoretical Physics, Cologne University, 
D-50923
K\"oln, Euroland}

\bigskip
* now back at  U.S.T.O., D\'{e}partement de Physique, L.E.P.M., 
B.P.1505
 El M'Naouar, Oran, Algeria.

\begin{abstract}
This third part of the lecture series deals with the question: Who will
pay for
your retirement? For Western Europe the answer may be ``nobody'', but 
for
Algeria the demography looks more promising.

\end{abstract}

\begin{figure}
\begin{center}
\includegraphics[angle=-90,scale=0.5]{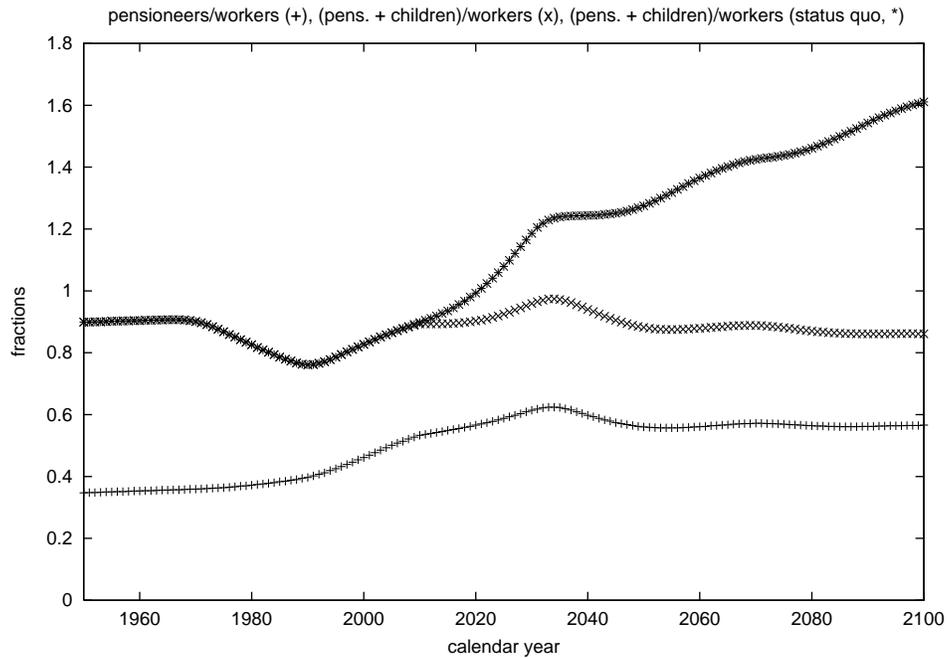}
\end{center}
\caption{Ratio of number of pensioneers to number of working age people (+)
and ratio of number of pensioneers plus number of children to working age people
(x,*). The curves (+,x) take into account a net immigration of 0.38 percent per
year and an increase of the average retirement age by half of the previous
increase  of the life expectancy. (Western Europe)
}
\end{figure}

\section{Introduction }
During the last two centuries in peaceful rich countries, people lived on
average longer and longer, while during the last few decades the number of
chidren born per women during her lifetime has sunken below the replacement
rate of slightly above 2. Also in many poorer countries the number of births
has fallen and the life expectancy increased. Thus the fear of overpopulation
of our planet Earth has to be modified by fear of old-age poverty: In the
year 2030 only those goods and services can be consumed by retired people 
which have
been produced by working-age people. A million dollars of old-age savings
can be halved by a ten-percent inflation rate over seven years, if not 
enough young people help me to live. This Econosociobiophysics problem is one 
of demography, not of money.

We present in an appendix details of the assumptions for our extrapolations into
the future. In the next section we deal with conditions as are typical for
Western Europe, to be followed by a section on the different problems of
Algeria. More literature on ageing models, including one applied to our 
demography \cite{cebrat}, is given in \cite{vancouver}.

\section{Western Europe}

Around 1970, the contraceptive pill reduced in the then two German states
the average number of babies born by a women during her lifetime below the
replacement level of two, to about 1.4. Spain and Italy followed later but
levelled at a lower plateau, while in France the number is higher, about
1.7. Life expectancy rises further though slower than during the first half of
the 20th century. Thus if people retire at an age of about 62 years, and if
around 2030 the strongest age cohort in Germany are the 70-year olds, problems
lie ahead. Only in recent years were they discussed in general newspapers.
As in science in general, we need open publications of extrapolation methods
and results. Only if many different simulations are compared can we see to what
extent they agree and thus may be relied upon.

The top curve in Fig.1 shows what happens if nothing is done: The average
retirement age is 62 years, and immigration and emigration cancel each other.
Then \cite{bomsdorf,stauffer,martins} the number of old people to be supported
by working-age people will increase drastically, while the total population
will decrease. We added here the number of children (up to age 20) to the
pensioneers since both groups are not fully ''working'' in the usual sense.
For the middle curve we assumed a net immigration of 0.38 percent per year,
starting now, and an increase of the retirement age by about half of the
increase of the life expectancy. Thus for every year which medical progress
gives us, about six month are given like a tax to the labour market, while the 
other six months are
leasure time after retirement. Now the ratio and the population are more stable.
If we do not count in the latter simulation the children (bottom curve), then 
the ratio of
pensioneers only to working age people is lower \cite{martins}. However, the
reduction of the expenses for children is mainly an effect of the past, not of
the future.

\begin{figure}
\begin{center}
\includegraphics[angle=-90,scale=0.5]{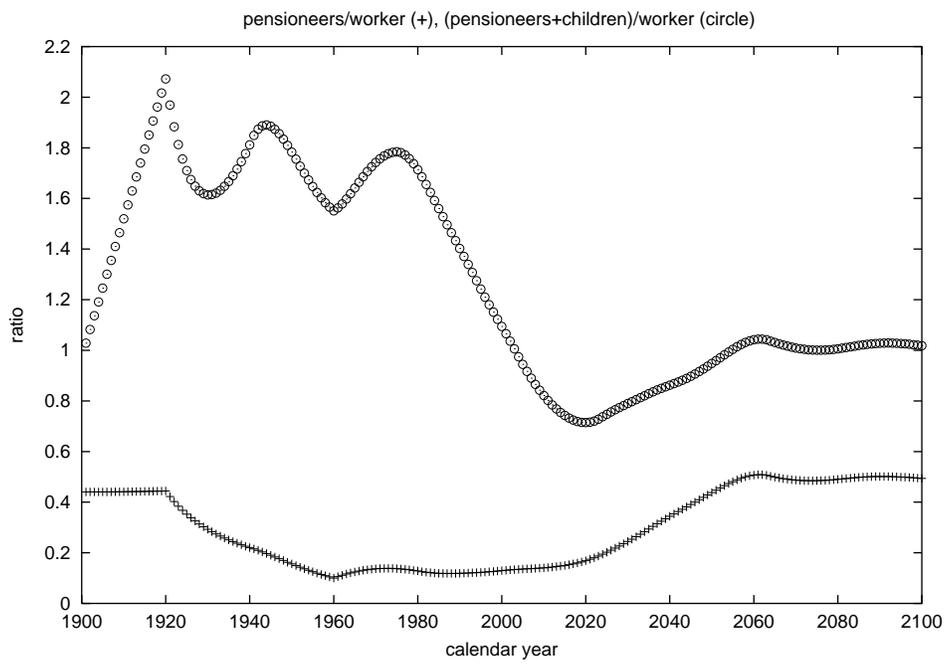}
\end{center}
\caption{Ratio of number of pensioneers to number of working age people (+)
and ratio of number of pensioneers plus number of children to working age
people (x). (Algeria)
}
\end{figure}

\begin{figure}
\begin{center}
\includegraphics[angle=-90,scale=0.5]{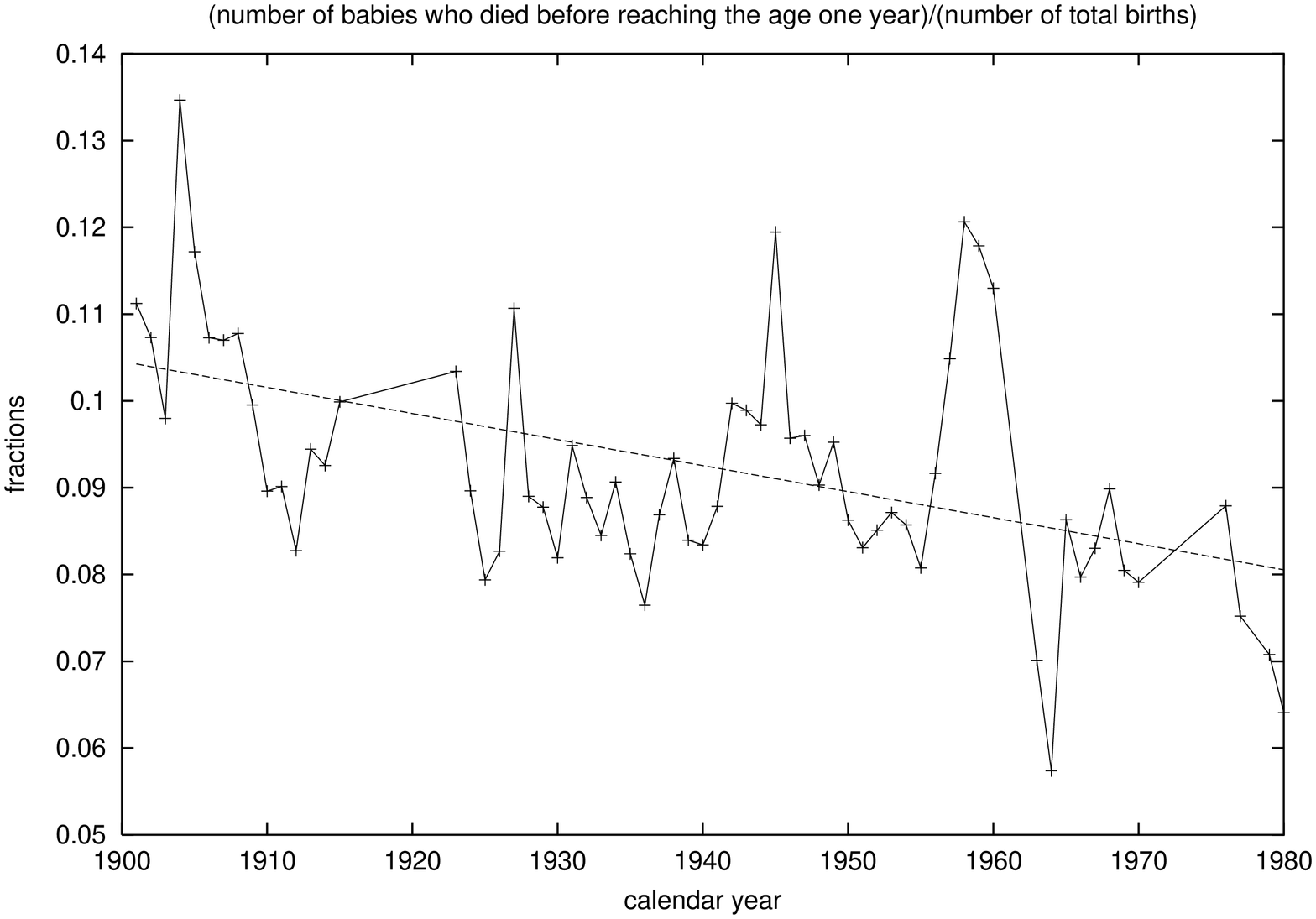}
\end{center}
\caption{Ratio of number of babies died before reaching the age one year
to number of total birth(+) (data of the National Office of the Statistics
ONS Algeria)
the fit line shows fractions increase of about 20 percent from 1980 until
1901.
}
\end{figure}

\section{Algeria}
During the first half of the previous century, the fertility was very 
large in North Africa compared to Europe. It reached the value 8.1 during 
the seventies in Algeria because of a low
average age of marriage in this country. Thirty years later, the average 
number of births per women (during her lifetime) becomes close to 2, whereas in 
France the fertility needed two centuries to pass from 6 in the middle of the 
17th century to 2 in the 1930's.  Algerian people are thus young. 

Figure 2 shows that the
number of children (up to age $20$) added to the pensioneers
(the retirement age in Algeria is $60$ years) obliged  workers to
support about two times their number until the year $2000$. Sixty years
afterwards the population will be older but the fractions remain constant (no 
fear of increasing). We assumed in Fig.2 the Gompertz slope $b$ (see appendix) 
to increase with time from $0.07$ in 1901 to $0.082$ in 1971 and to remain 
constant thereafter. Only fertility data from 1950 on is available in Algeria.
The fertility is constant with a mean value of $7.3$ from $1950$ to $1980$ and
then decreases abruptly til 2004 to reach a value $2.04$; it is assumed to 
stay constant at this value thereafter. The sixty years period necessary to 
reach the steady state, corresponds to the age of retirement. In figure 3, we 
show that the ratio of the number of babies dying in their first year to the 
total number of births decreased by about $20$ percent from 1901 to 1980. Thus, 
we made a correction on the fertility data (in fig.2) by reducing them by the
number of children dying before they reach maturity.
We noticed also that the 
greatest emigration rate of Algerian people was between 1950 and 1970 but 
remains weak compared to the rate of births and does not influence the 
population evolution. In our simulation 
we then neglected the emigration in such calculations. However, this simulation 
did not account the rate of unemployeds  
which was very small during the period of socialism but reaches now 17 percent 
of population. However, the main prediction of Fig. 2 is an increase of the 
social load for old age by 400 percent starting from 2020, while that for 
children and old age combined will stabilize at the level around the year 2000.

\section{Summary}

With rising life expectencies and falling births, the demographic problems of
rich countries can be alleviated by controlled immigration and a moderate
increase of the average retirement age. That policy requires that first the
unemployment is reduced appreciably. For Algeria, on the other hand, emigration
could not affect sensitively the evolution of pensioneers, but their rate should
be multiplied by a factor four after 15 years from now on which would create 
a real economic problem were it not offset by a reduction of the number of 
children.

\bigskip
%\begin{theacknowledgments}
LZ thanks the DAAD for supporting a one-year part of his thesis work in
Cologne.
We thank W.J. Paul for suggesting to add the children to the 
pensioneers.

\section{Appendix}

According to the Azbel lectures at this seminar, in all different countries and
centuries, the probability of humans to survive up to a fixed age is a
universal function of the life expectancy; we do not have to apply this
universality to yeast cells for the purpose of human demography. Thus we use
Germany as typical Western European country, without taking into account the
effects of World War II.

The mortality function $\mu = - d \ln S(a)/da$, where $S(a)$ is the number of
survivors from birth to age $a$, is assumed to follow a Gompertz law for adults:
$m \propto b\exp[(a-X)b]$ since the deviations at young age occur at such low
mortalities that they are not relevant if we want to be accurate within a few
percent. The deviations at old age \cite{robine} are not yet reliably
established and may also be negligible as long as the fraction of centenarians
among pensioneers is very small.

The Gompertz slope $b$ was assumed to increase linearly with time from 0.07 in
1821 to 0.093 in 1971 and to stay constant thereafter, in contrast to Bomsdorf
\cite{bomsdorf} and Azbel \cite{azbel} but in agreement with Yashin et al
\cite{yashin}; see also Wilmoth et al \cite{wilmoth}. Instead, the
characteristic age $X$ was constant at 103 years until 1971 and then increased
each year by 0.15 years to give a rising life expectancy.
Also these deviations from universality are not yet established reliably.
(Therefore we ignored the effect for Algeria, keeping $X=103$ constant there.)

Babies are born by mothers of age 21 to 40 with age-independent probability.
The average number of children born per women over her lifetime and reaching
adult age is assumed to be $2.2 - 0.4\tanh[(t-1971)/3]$ recently.
Immigrants are assumed to be 6 to 40 years old with equal probability, and their
number per year equals a fraction $c = 0.38 \%$ of the population, adjusted to
give a constant total population.

After the year 2010, the retirement age is increased by 60 percent of the
increase of life expectancy at birth to 73 in 2100 at a life expectancy then
of 99 years; for the problem year 2030 these ages are 64 and 84 years.

The program is available from stauffer@thp.uni-koeln.de as rente16.f.


\begin{thebibliography}{99}
\bibitem{cebrat} A. {\L}aszkiewicz, Sz. Szymczak, S. Cebrat, Int. J.
Mod. Phys. C 14, 1355 (2003); S. Cebrat and A. {\L}aszkiewicz. J. 
Insur.
Medicine 37, 3 (2005).
\bibitem{vancouver} D. Stauffer, p. 131 in: Thinking in
     Patterns, ed. by M.M. Novak, World Scientific, Singapore 2004.
For an older but much more detailed review see S. Moss de Oliveira, 
P.M.C.
de Oliveira and D. Stauffer, {\it Evolution, Money, War and Computers},
Teubner, Stuttgart and Leipzig 1999.
\bibitem{bomsdorf} E. Bomsdorf, Exp. Gerontology 39, 159 (2004).
\bibitem{stauffer} D. Stauffer, Exp. Gerontology, 37, 1131 (2002).
\bibitem{martins} J.S.  S\'a Martins, D. Stauffer, Ingenierias (Univ.
Nuevo Leon, Mexico)  7, No.22, p.35 (Jan-Mar 2004).
\bibitem{robine} J.-M. Robine, J.W. Vaupel, Exp. Gerontology 36, 915
(2001); K. Suematsu, J. Theor. Biol. 201, 231 (1999).
\bibitem{azbel} M. Ya. Azbel, Proc. Roy. Soc. B 263, 1449 (1996).
\bibitem{yashin} A.I. Yashin, A.S. Begun, S.I. Boiko, S.V.Ukraintseva, 
J.
Oeppen,
Exp. Gerontology 37, 157 (2001).
\bibitem{wilmoth} J.R. Wilmoth, L.J. Deegan, H.Lundstr\"om, S. 
Horiuchi:
Science 289, 2366 (2000).

\end{thebibliography}
\end{document}